\documentclass[12pt,preprint]{aastex}

\newcommand\TA{\tablenotemark{a}}
\newcommand\TB{\tablenotemark{b}}
\newcommand\TC{\tablenotemark{c}}

\newcommand{\Hiirs}{\ion{H}{2} regions}

\newcommand{\mcnd}{\multicolumn{2}{c}{\nodata}}
\newcommand{\xion}{(ion)}

\shorttitle{The ADF in \ion{H}{2} regions}
\shortauthors{Peimbert, \& Peimbert}
\received{}

\begin{document}

\title{Densities, temperatures, pressures, and abundances derived from \ion{O}{2} 
recombination lines in \ion{H}{2} regions and their implications}

\author {Antonio Peimbert\footnotemark[1]}
\email{antonio@astro.unam.mx}

\and 

\author{Manuel Peimbert\footnotemark[1]}
\email{peimbert@astro.unam.mx}

\footnotetext[1]{Instituto de Astronom\'ia, Universidad Nacional
  Aut\'onoma de M\'exico, Apdo. Postal 70-264, M\'exico 04510 D.F.,
  Mexico}

\begin{abstract}
Based on high quality observations of multiplet V1 of \ion{O}{2} and
the NLTE atomic computations of \ion{O}{2} we study the density and
temperature of a sample of \ion{H}{2} regions. We find that the 
signature for oxygen rich clumps of high density and low temperature 
is absent in all objects of our sample: one extragalactic and eight 
{Galactic} \ion{H}{2} regions. The temperatures derived from: 
a) recombination lines of \ion{O}{2}, and b) {recombination lines of} \ion{H}{1}
together with Balmer continua are lower than those derived from
forbidden lines, while the densities derived from recombination lines
of \ion{O}{2} are similar or smaller than densities derived from
forbidden lines. Electron pressures derived from collisionally excited
lines are about two times larger than those derived from
recombination lines. These results imply that the proper abundances
are those derived from recombination lines and suggest that other 
processes in addition to direct photoionization, such as dissipation 
of turbulent energy in shocks, magnetic reconnection, and shadowed 
regions, might be responsible for the large ADF and $t^2$ values 
observed in \ion{H}{2} regions.
\end {abstract}


\section{Introduction}\label{Sintro}

Since the collisional intensities of O lines, as well as other heavy
elements, are several orders of magnitude stronger than the 
recombination lines, most abundances of heavy elements relative to 
those of hydrogen have been derived from collisionally excited lines. 
In addition usually collisional abundances are only derived under the 
assumption of chemical homogeneity and constant temperature.

O/H abundance ratios derived from recombination lines 
of O and H are higher than those derived from the ratio of a  
collisionally excited line (CL) of oxygen to a recombination line 
(RL) of H, this effect is called the abundance discrepancy problem, 
and the ratio of both types of abundances is called the abundance 
discrepancy factor (ADF). This problem also applies to other heavy
elements like C, N, and Ne. 

There are several explanations for the ADFs present in the literature
for example: temperature variations in a homogeneous medium,
inhomogeneous chemical composition, errors in the atomic 
parameters, and overestimation of the intensity of weak 
recombination lines. Errors in the atomic parameters have been ruled 
out because the ADF values vary from object to object, and the 
overestimation of the intensity of the weak lines has also been ruled 
out because the ADF problem persists for objects where weak 
unblended RLs have been measured with a S/N higher than 10.

In a chemically homogeneous medium  $I$(O,RL)/$I$(H,RL) is 
proportional to the O/H ratio and is almost independent of the 
electron temperature. Alternatively $I$(O,CL)/$I$(H,RL) does depend 
on the electron temperature in such a way that in the presence of temperature 
variations the O/H abundances derived from temperature 
determinations based on CLs, assuming constant temperature, yield 
abundances smaller than the real ones giving rise to the presence of
an ADF.

In a chemically inhomogeneous medium CLs are expected to 
originate mainly in regions that are relatively metal-poor, 
temperature-high and density-low, while the RLs are expected to 
originate mainly in regions that are relatively metal-rich, 
temperature-low, and density-high.

It is the purpose of this paper to study the cause of the ADF values.
We will concentrate on the O/H ADF in \ion{H}{2} regions 
considering two options: a) the presence of temperature variations in 
a chemically homogeneous medium and b) the presence of chemical 
inhomogeneities. It should be mentioned that chemical 
inhomogeneities also produce temperature variations. In addition to 
the evidence in favor of the presence of temperature variations based 
on chemical abundance determinations there is evidence of 
temperature variations based on high spatial resolution observations 
of the Orion nebula and the Ring nebula \citep{ode03,ode13}.

The data of \ion{H}{2} regions obtained with the Ultraviolet Visual 
Echelle Spectrograph, UVES, and the very large Kueyen telescope in 
Chile, VLT are specially suited for the study of faint emission lines 
due to their high quality, produced by their high spectral 
resolution and their high S/N. There are nine \ion{H}{2} regions 
that have been observed with this equipment 
\citep{pea03,est04,gar04,gar05,gar06,gar07}. The observational data 
already published will be used in this paper. Hereafter we will refer 
to this data set as the UVES set. These papers  show fractions of the 
spectra presenting the region of multiplet V1 of \ion{O}{2} where the 
spectral resolution and the S/N can be appreciated. With the 
exception of the densities and temperatures derived in this paper 
and those derived by \citet{mna13} from the UVES set, all the other 
values and densities presented in this paper were computed in the 
original papers that contain the observational UVES set.

In Section \ref{Stemp} we derive the electron temperatures from the 
ratio of [\ion{O}{3}] lines, $T$(4363/4959),  and from the ratio of the
sum of the 8 recombination lines of multiplet V1 of \ion{O}{2} to the 
4959[\ion{O}{3}] line, $T$(V1/4959), and from these values the 
average temperature and mean square temperature variation of the 
O$^{++}$ region, $T_0$(\ion{O}{2}) and $t^2$(\ion{O}{2}). We also 
compare these $T_0$ and $t^2$ values with those derived from 
other forbidden and permitted line ratios, such as those derived from 
the Balmer lines and continuum. In Section \ref{Sden} we derive 
electron densities based on the \ion{O}{2} recombination lines and 
compare them with those derived from forbidden lines. In Section 
\ref{Spressure} we compare the pressure derived from collisionally 
excited lines with that derived from recombination lines. In Section 
\ref{Sdoradus} we compare the O/H ratio derived with eight different 
methods for 30 Doradus, an \ion{H}{2} region in the Large Magellanic 
Cloud. The discussion and conclusions are presented in Sections 
\ref{Sdisc} and \ref{Sconc} {respectively}.

\section{Temperature determinations based on the \ion{O}{2} and 
[\ion{O}{3}] lines}\label{Stemp}

Earlier calculations of the effective recombination coefficients of
\ion{O}{2} were derived under the assumption that the
fine-structure levels of the ground term of the recombining ion are
thermally populated in proportion to their statistical weights
\citep{pem93,sto94}, an assumption that has been shown to be
inaccurate under low density nebular conditions \citep{rui03,pea05}.

New ab initio calculations of the effective recombination coefficients, 
valid down to very low temperatures and taking into account the 
density dependence of the level populations of the ground states of 
the recombining ion, are now available for the recombination 
spectrum of \ion{O}{2} \citep[][Storey unpublished]{bas06,liu12,fan13}.

When choosing which \ion{O}{2} RLs to use to derive the 
temperature and the abundace of the O$^{++}$ region there 
are three strong reasons to use the 
sum intensities of the eight lines of the V1 multiplet, $I$(V1): a) V1 
is the brightest multiplet in the visual region, b) the intensity of a 
single line of the V1 multiplet is density dependent and the error in 
the density determination propagates into the resulting temperature 
or abundance, while the total intensity of the multiplet is practically 
density independent, and c) the error in the determination of the 
intensity of the whole multiplet, if the resolution is high enough to 
detect all the lines without blends from other ions, is considerably 
smaller than that of a single line.

\begin{figure}
\begin{center}
\includegraphics[angle=0,scale=0.82]{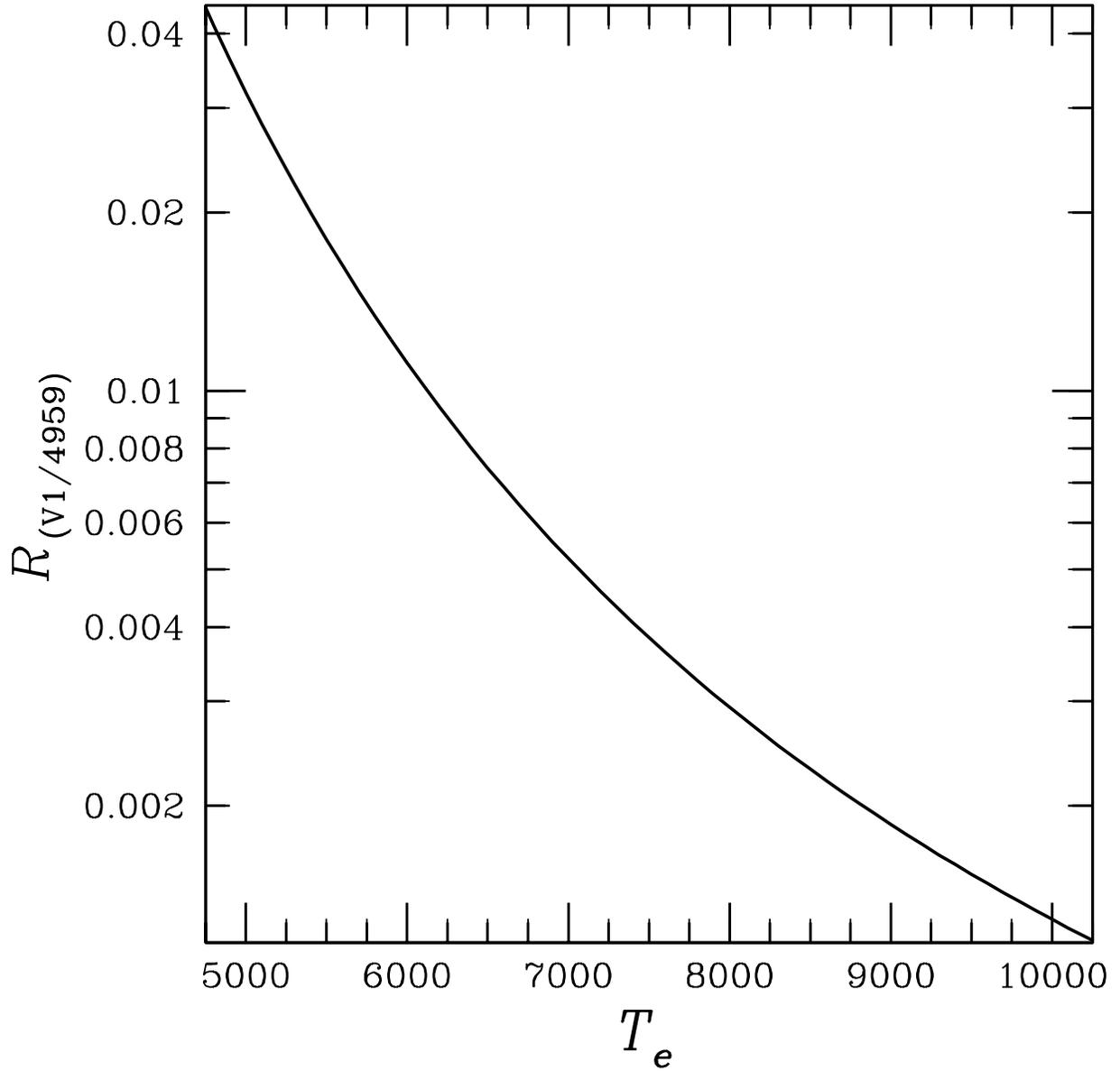}
\end{center}
\caption{$T$ versus $R_{({\rm V1}/4959)} = \varepsilon_{\rm V1}  / \varepsilon_{4959}$
derived from equation \ref{Ere-V1/4959}. For all the objects in the UVES sample 
$T_e(V1/4959)$ is smaller than $T_e(4363/4959)$, showing the prescence 
of temperature inhomogeneities.
\label{Fv1-T}}
\end{figure}

The emission coefficients per ion per electron per cm$^{-3}$ for 
these lines: $\varepsilon_{\rm V1}$, $\varepsilon_{4959}$, and 
$\varepsilon_{4363}$ as a function of temperature in the 5000 to 
15000K range are given by:
\begin{equation}
\label{Ee-V1}
\varepsilon_{\rm V1}(T_e)= C_{\rm V1} T_e^{-0.755},
\end{equation}
\begin{equation}
\label{Ee-4959}
\varepsilon_{4959}(T_e)= C_{4959} T_e^{-0.34}exp (-29160/T_e),  
\end{equation}
and
\begin{equation}
\label{Ee-4363}
\varepsilon_{4363}(T_e)= C_{4363} T_e^{-0.34}exp (-62120/T_e),
\end{equation}
\citep{len94,men99} where the C values are constants that depend 
on atomic parameters. The observed intensity of each line is given 
by:
\begin{equation}
\label{EI-e}
I= \int{ \varepsilon(T_e) n_e n({\rm O^{++}})\over r^2} dV,
\end{equation}
where $V$ is the emitting volume and $r$ is the distance to the 
source. If we assume that $T_e$ is constant in the observed volue 
we have that
\begin{equation}
\label{EI-e-W}
I=\varepsilon(T_e) \int{n_e n({\rm O^{++}})\over r^2} dV =
\varepsilon(T_e) W({\rm O^{++}}),
\end{equation}
where $W({\rm O^{++}})$ is closely related to the emission measure 
and is common to all O$^{++}$ lines; therefore the ratio of two line 
intensities only depends on the {emissivities}, consequently
\begin{equation}
\label{Ere-V1/4959}
{I({\rm V1}) \over I(4959)} =
{\varepsilon_{\rm V1}(T_e) \over \varepsilon_{4959}(T_e)} =
R_{({\rm V1}/4959)}(T_e)=
{C_{\rm V1} \over C_{4959}}T_e^{-0.415}exp (29160/T_e),
\end{equation}
and
\begin{equation}
\label{Ere-4363/4959}
{I(4363) \over I(4959)} =
{\varepsilon_{4363}(T_e) \over \varepsilon_{4959}(T_e)} =
R_{(4363/4959)}(T_e)= 
{C_{4363} \over C_{4959}}exp (-32940/T_e),
\end{equation}
where $R_{({\rm V1}/4959)}$ and $R_{(4363/4959)}$ are the 
ratios of the emission coefficients; and 
$C_{\rm V1}/C_{4959} = 6.56 \times 10^{-5}$ and 
$C_{4363}/C_{4959} = 0.496$ \citep{sto94,len94,men99}. In 
Figure \ref{Fv1-T} we present Equation \ref{Ere-V1/4959}, the relation between 
$R_{({\rm V1}/4959)}$ and $T$; equation \ref{Ere-4363/4959} is the 
most often used to determine the temperature of photoionized 
regions.

From equations \ref{Ere-V1/4959} and \ref{Ere-4363/4959} we 
have determined $T_e({\rm V1}/4959)$ and $T_e(4363/4959)$, 
which, toghether with the $T$(\ion{O}{2}) values derived by 
\citet{mna13}, are presented in Table \ref{Tt-OII}. For all 
objects of the UVES set we find that $T_e(4363/4959)$ is higher 
than $T_e({\rm V1}/4959)$, considering that the dependence {on} 
$T_e$ is {stronger} for $I(4363/4959)$ than for $I({\rm V1}/4959)$, 
it is possible that the difference between the two sets of 
temperatures could be due to the presence of temperature variations 
over the observed volume. 

We decided to follow the formalism introduced by \citet{pem67} 
to determine the basic parameters of the temperature structure, 
$T_0(\rm{O^{++}})$ and $t^2(\rm{O^{++}})$, where 
\begin{equation}
\label{Edef-T0}
T_0({\rm O^{++}})=
\frac{\int T_e n_e n({\rm O^{++}})dV}
{\int n_e n({\rm O^{++}})dV},  
\end{equation}
and
\begin{equation}
\label{Edef-t2}
t^2({\rm O^{++}})=
\frac{\int (T_e - T_0({\rm O^{++}}))^2 n_e n({\rm O^{++}})dV}
{T_0({\rm O^{++}})^2 \int n_e n({\rm O^{++}})dV}.  
\end{equation}

In the presence of temperature variations it is not possible to 
simplify equation \ref{EI-e} in the way is done in equation 
\ref{EI-e-W}. In appendix \ref{appendix} we present a way to relate 
line intensity ratios to the temperature structure. In this formalism we 
need two independent line intensity ratios (of lines of the 
same ion and with little density dependence) to derive 
$T_0({\rm O^{++}})$ and $t^2({\rm O^{++}})$.

From equation \ref{E-T-t2} and the temperature dependence of 
$\varepsilon_{\rm V1}$, $\varepsilon_{4959}$, and 
$\varepsilon_{4363}$ (equations \ref{Ee-V1}, \ref{Ee-4959}, and 
\ref{Ee-4363}) we can write $T_e(V1/4959)$ and $T_e(4363/4959)$ as a
function of $T_0$ and $t^2$:
\begin{equation}
\label{ET0-OIII}
T_e(4363/4959) = T_0({\rm O^{++}})
\left[ 1 + \left( \frac{91300}{T_0({\rm O^{++}})} - 2.68\right)
\frac{t^2({\rm O^{++}})}{2} \right],
\end{equation}
and
\begin{equation}
\label{ET0-OII}
T_e(V1/4959) = T_0({\rm O^{++}})
\left[ 1+\left( {29160 \over T_0({\rm O^{++}})} - 3.095 + 
{ 0.415 \over {29160 \over T_0({\rm O^{++}})} +0.415} \right) {t^2({\rm O^{++}}) \over 2} \right].
\end{equation}
Therefore from equations \ref{ET0-OIII} and \ref{ET0-OII} 
we have derived $T_0$(O$^{++}$) and $t^2$(O$^{++}$), 
presented in Tables \ref{Tt-OII} and \ref{Tt-var}.

{\citet{fan13} and \citet{mna13} have used the 
$I$(4649)/$I$(4089) \ion{O}{2} ratio to derive the electron 
temperature of the O$^{++}$ zone using only recombination lines
(where $\lambda$ 4649 belongs to the V1 multiplet, and  
$\lambda$ 4089 belongs to the V48a multiplet).} We have decided not 
to use this ratio for the following reasons: a) $\lambda$ 
4089 has been detected only in three of the nine \ion{H}{2} regions
in our sample: 30 Doradus, Orion, and NGC 3576, while for the 
other six regions only an upper limit to the intensity of the 
$\lambda$ 4089 line can be obtained, that corresponds to a lower 
limit in the temperature, b) the $I$(4649)/$I$(4089) ratio depends 
very weakly on the electron temperature and in the three regions 
where it has been detected the error in the ratio is in the 15\% 
to 20\% range, an error of 15\% in the $ I$(4649/4089) \ion{O}{2} 
ratio implies an error of about 3500 K for a temperature of 8000 K, 
c) $\lambda$4089 can have a significant contribution due to the 
\ion{Si}{4} line at $\lambda$4088.86, in this case only a lower limit 
of the temperature can be obtained from the $I$(4649)/$I$(4089) 
ratio.

\begin{figure}
\begin{center}
\includegraphics[angle=270,scale=0.62]{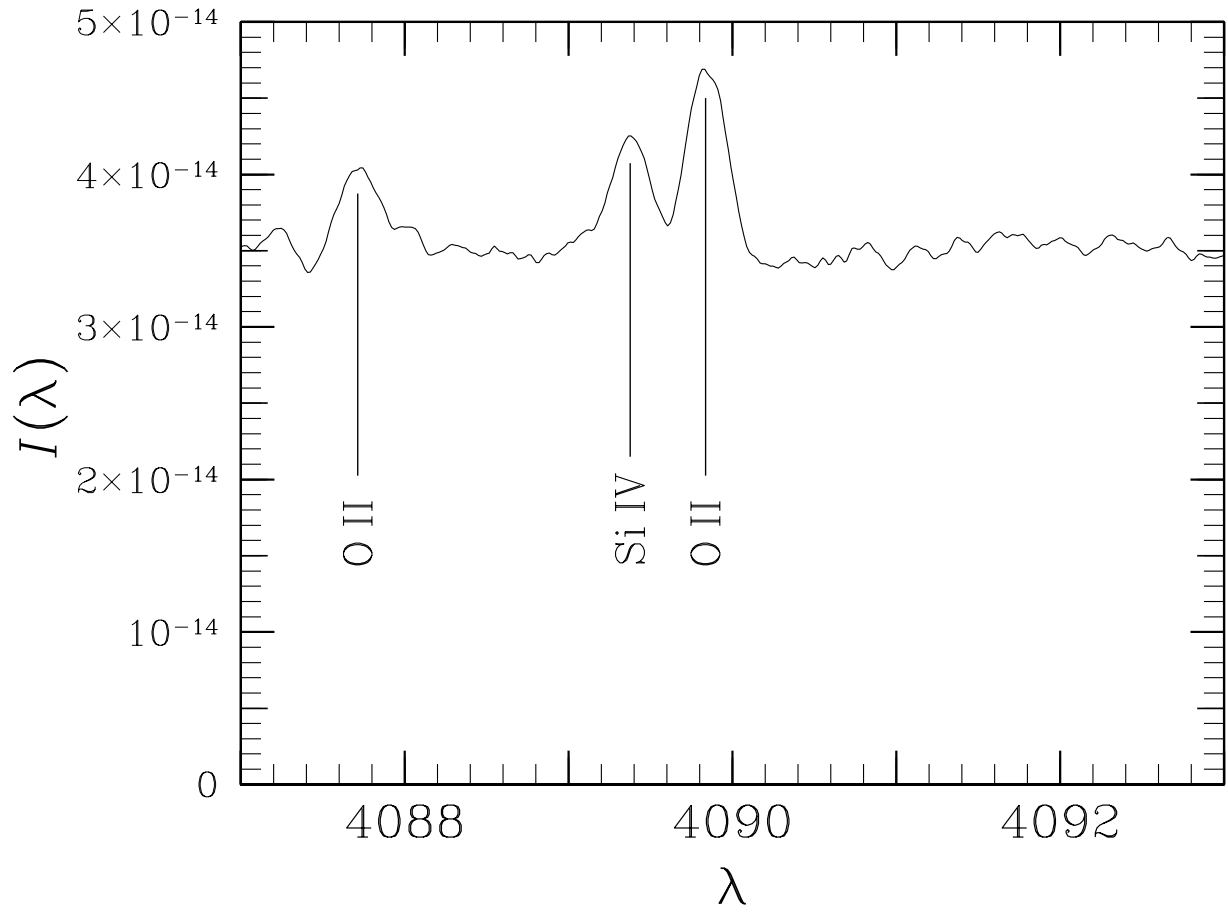}
\end{center}
\caption{$I(\lambda)$ versus $\lambda$. Section of the UVES spectrum of the Orion
nebula by \citet{est04} showing the presence of the \ion{Si}{4} $\lambda$4088.86 
line in emission.
\label{FSiIV}}
\end{figure}

There is evidence in favor of a contribution to the $\lambda$4089 
feature due to the presence of the \ion{Si}{4} line at 
$\lambda$4088.86 for two of the three regions where 
$\lambda$4089 has been detected. For the Orion nebula a 
line around $\lambda$4116.10 has been 
detected and \citet{est04} suggested that it might be due to a line 
of the v2F0-6D multiplet of Fe II] at 4116.067, we do not agree
with this suggestion because the other 5 lines of multiplet 
v2F0-6D were not detected: $\lambda\lambda$4030.970, 
4065.317, 4131.621, 4184.051 and 4243.085. We suggest that 
the line at $\lambda$4116.10 is the weaker one of the doublet 
of \ion{Si}{4} that includes  $\lambda$4088.86. To confirm this 
suggestion we looked again at the original UVES spectrum of the 
Orion nebula (see Figure \ref{FSiIV}) and found that indeed the 
$\lambda$4088.86 line is present with an intensity of 0.017 after 
correcting for reddening, where $I$(H$\beta$) = 100. Therefore the 
observed $I$(4088.86)/$I$(4116.10) ratio is equal to 2.4 in good 
agreement with the theoretical ratio that amounts to two.
 
Furthermore the presented $\lambda$$\lambda$4088.86, 4089.29, 
and 4116.10 line intensities are lower limits to the real intensities 
because in Orion there is a substantial component of the continuum 
due to dust scattered light \citep{ode65} that is expected to show
the \ion{Si}{4} lines in absorption. The dust scattered light is mainly 
due to the brightest stars in the Trapezium with B0.5V, B0V, O7V 
and O9.5V spectral types, for components A, B, C, and D, 
respectively \citep{iri65,hof82,con71}, the \ion{Si}{4} lines reach 
their peak intensities around the spectral type O9.5 to B0.5
\citep{con73,rud36}. \ion{Si}{4} lines in absorption have been detected in 
component C of the Trapezium see Figure 4 of \citet{est98} . We 
have not estimated the correction due to the underlying absorption 
that affects the \ion{Si}{4} $\lambda$4088.86 and the \ion{O}{2} 
$\lambda$4089.29 line intensities in emission. 

For 30 Doradus, where the feature at $\lambda$4089 has also been 
detected, there is an additional argument in favor of the presence of 
the \ion{Si}{4} line based on the central wavelength of the observed 
feature. The theoretical displacement of the \ion{Si}{4} 
$\lambda$4088.86 line is 0.43 \AA\ to the blue of the \ion{O}{2} line 
at $\lambda$4089.29.  For 30 Doradus the observed feature 
identified as $\lambda$4089.29 is shifted towards the blue by 
0.31 \AA\ relative to the wavelength frame defined by the \ion{O}{2} 
lines at $\lambda$$\lambda$4072.16, and 4078.84, the shift 
suggests that a substantial fraction of the blend could be due to 
\ion{Si}{4}. By assuming that one fifth to two fifths of the 
$\lambda$4089 blend is due to \ion{Si}{4} we obtain a temperature 
in the 5000 to 12600 K range, while by assuming that the 
$\lambda$4089 intensity is due only to \ion{O}{2} a temperature 
of 400 K is obtained. 
 
Another argument in favor of the presence of \ion{Si}{4} in gaseous 
nebulae is that the weaker line of the  \ion{Si}{4} doublet at 
$\lambda$4116.10 has been detected in some planetary nebulae 
of intermediate degree of ionization like NGC~6543, NGC~6572, 
IC~4997 and of higher degree of ionization like NGC~7009 and 
NGC~7662 \citep{hyu00,hyu94a,hyu94b,hyu95,all66}. 

For the other six \ion{H}{2} regions of our sample $\lambda$4089 
was not detected, therefore an estimate of the $\lambda$4089 
intensity is only an upper limit of the \ion{O}{2} feature and 
consequently the $T_e$ values derived from Figure 2 of 
\citet{mna13}(that shows $I$(4649)/$I$(4662) and 
$I$(4649)/$I$(4089) as a function of density and temperature) 
become only lower limits to the real $T$(\ion{O}{2}) value.

From the $T_0({\rm O}^{++})$ values presented in Table \ref{Tt-OII} 
we have estimated that, to determine the $\lambda$4089 line 
intensity with an error of 15\% for the six objects where it was not 
detected, we need new observations with signal to noise ratios 
4 to 13 times better than those present in the UVES set (4 for M8 
and 13 for M20).

In Table \ref{Tt-OII} we also present the temperatures derived by 
\citet{mna13}, from the $I(4649)/I(4089)$ ratio, from the 
arguments and results presented above we consider that their 
results are only lower limits to the electron temperature, with the 
exception of the Orion nebula value that might be an upper limit to 
the temperature due to the contribution of dust scattered light 
showing the \ion{Si}{4} line in absorption.

To study the possibility of chemical inhomogeneities we present in 
Tables \ref{Tt-lit} and \ref{Tt-var} temperatures and mean square 
temperature variations derived from the \ion{O}{2} and [\ion{O}{3}] 
lines by us in this paper, and from the \ion{He}{1} and \ion{H}{1} 
lines as well as the Balmer continuum from the UVES set in the 
literature. From {these} tables we find similar values for the O, He, and 
H temperatures and mean square temperature variations in 
agreement with chemical homogeneity for this group of \ion{H}{2} 
regions.

\section{Density determinations based on the \ion{O}{2} lines}\label{Sden}

\begin{figure}
\begin{center}
\includegraphics[angle=0,scale=0.82]{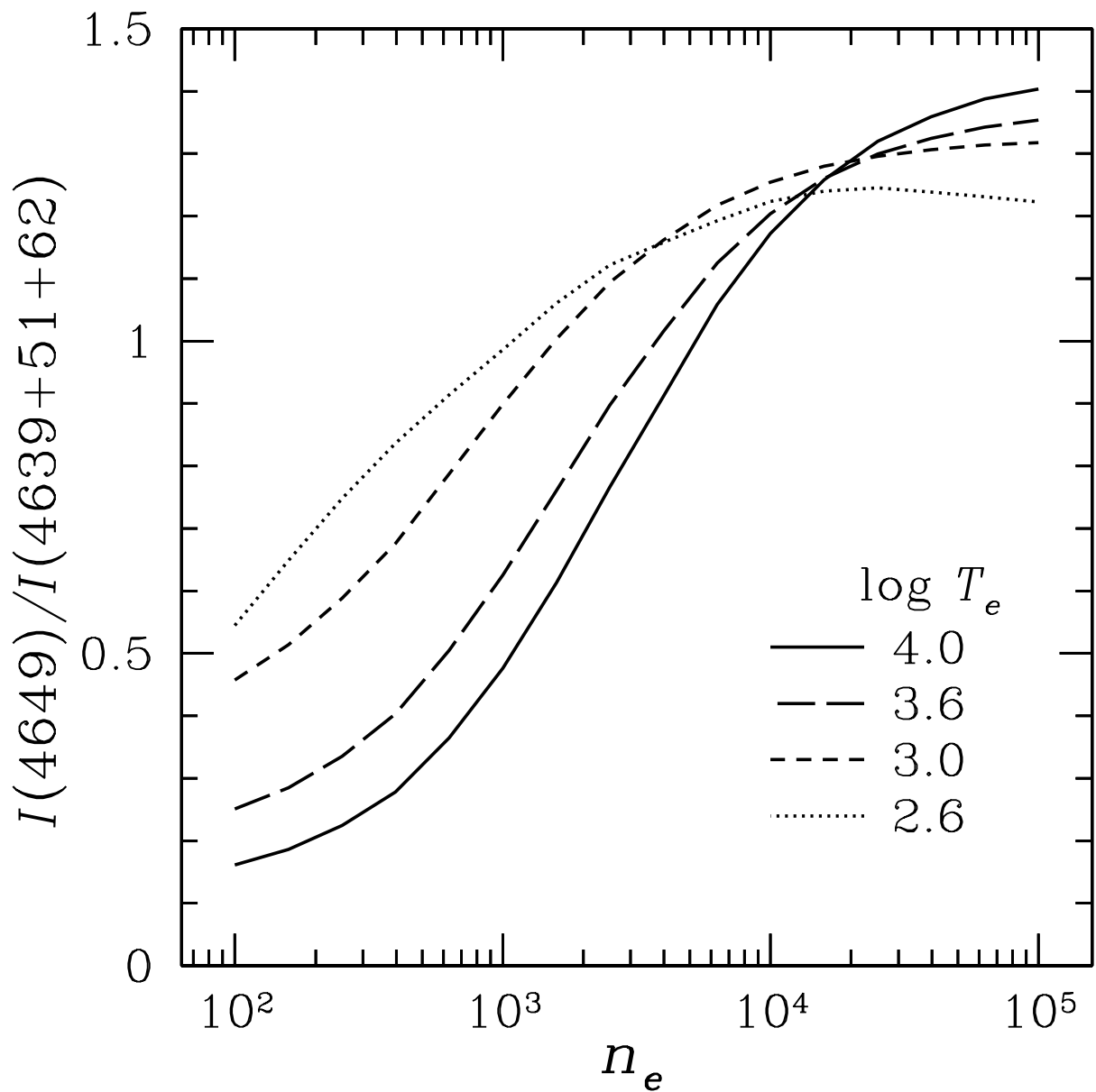}
\end{center}
\caption{$I$(4649)/$I$(4639 + 4651 + 4662) versus $n_e$(\ion{O}{2}). Intensity density
diagram of \ion{O}{2} for different temperatures derived from the unpublished computations by
Storey \citep{bas06,liu12,fan13}.
\label{Fne_OII}}
\end{figure}

The electron densities for the \ion{H}{2} regions were derived from Figure \ref{Fne_OII}, 
where we plot the predicted $I$(4649)/$I$(4639 + 4651 + 4662) ratio from the 
atomic data by Storey \citep[unpublished, see also] []{fan13} and are presented in Table \ref{Tdensities}. 
We decided to use this ratio because the error in the $I$(4639 + 4651 + 4662) 
value is smaller than the error in the $I$(4662) value and the behavior versus 
density of $I$(4639) and $I$(4651) is similar to that of $I$(4662). The use 
of Figure \ref{Fne_OII} requires a temperature, {and} we are using the $T_0$(\ion{O}{2})
temperature presented in Table \ref{Tt-OII}, derived 
from the $I$(4959)/$I$(V1) and  $I$(4959)/$I$(4363) ratios
that is considerably more accurate than the temperature derived from the
$I$(4649)/$I$(4089) ratio for the reasons presented in the previous section.

For M20 $I$(4639) was not measured, and the errors in the determination
of $I$(4649), $I$(4651), and $I$ (4662) are the largest of the sample and 
were not estimated, therefore we did not obtain the density for this object.

Also in Table \ref{Tdensities} we include the densities derived from CLs presented 
in the VLT UVES papers. The atomic data used to derive the \ion{O}{2} densities 
from RLs by \citet{mna13} and us is the same, the different results come 
from the different \ion{O}{2} lines used to determine the density
and because we used the $I$(4959)/$I$(V1) ratio to determine the temperature 
while McNabb et al. used other \ion{O}{2} lines to determine the 
temperature.

\section{Electron pressure in \ion{H}{2} regions}\label{Spressure}

{We determined the pressures using the ideal gas equation.} In Table 
\ref{Tpres} we present the ratio of the pressure derived from the 
collisionally excited lines to the pressure derived from the recombination 
lines, $P$(CLs)/$P$(RLs). For $P$(CLs) we adopted the following equation,
\begin{equation}
\label{Epres}
P({\rm CLs}) = n_e\left<{\rm CLs}\right>kT_e(4363/4959),
\end{equation}
where $n_e\left<{\rm CLs}\right>$, the average of the density 
determinations from CLs, was 
obtained from the original papers and is presented in Table 
\ref{Tdensities}, {$k$ is the Boltzmann constant,} and $T_e(4363/4959)$ is presented in Table 
\ref{Tt-OII}. The $P$(RLs) were obtained from $n_e$(\ion{O}{2}) and $T_0$(O$^{++}$). 
For comparison we also present the ratio of pressures derived from 
the results by \citet{mna13}, where the $P$(CLs) are the values derived by 
us and the $P$(RLs) are the values derived from the densities and temperatures
presented in Table \ref{Tt-lit} of \citet{mna13} (values also presented in the last
column of Tables \ref{Tt-OII} and \ref{Tdensities} {\rm of this paper}).

The pressure ratios derived from our data are in the 0.85 to 3.95 range 
with an average value of 2.4. We consider that these ratios are important 
clues to study the process or processes that are responsible for the 
temperature variations present in \ion{H}{2} regions.

\begin{figure}
\begin{center}
\includegraphics[angle=270,scale=0.62]{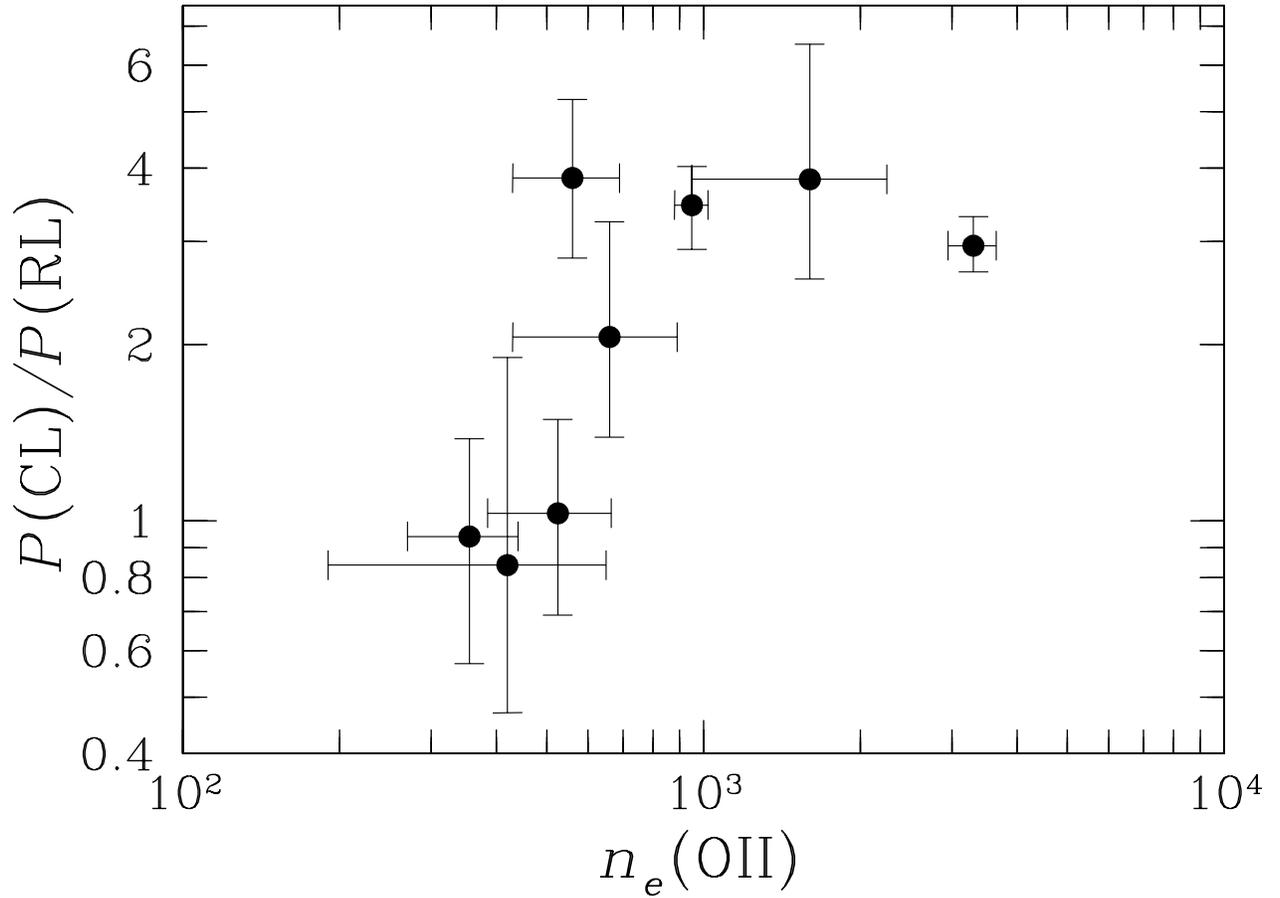}
\end{center}
\caption{$P$(CL)/$P$(RL) versus $n_e$(\ion{O}{2}). Ratio of pressures derived
from collisional to recombination lines as a function of the density obtained from
the \ion{O}{2} recombination lines. The data corresponds to the \ion{H}{2} regions
of the UVES set.
\label{Fpressure}}
\end{figure}

In Figure \ref{Fpressure} we show the $P$(CLs)/$P$(RLs) versus the $n_e$(\ion{O}{2})
values. This figure shows a trend of higher $P$(CLs)/$P$(RLs)
with higher $n_e$(\ion{O}{2}) values, the lowest density \ion{H}{2} regions 
show pressure ratios close to one, while the high density \ion{H}{2} regions 
show pressure ratios close to four. A seemingly stronger correlation could be presented 
by plotting $P$(CLs)/$P$(RLs) versus $n_e$[\ion{Cl}{3}], but it is not more 
meaningful because $n_e$[\ion{Cl}{3}] is a positive ingredient in the pressure ratio.

These correlations indicate that there is a mechanism capable of producing hot clumps
of high density. Since  $P$(CLs)/$P$(RLs) is generally higher than one, it
follows that hot, rather than cold, high density clumps are the dominant 
cause of temperature inhomogeneities{. Furthermore} this correlation shows that 
the mechanism that
produces these hot clumps is more efficient at higher densities than at lower
densities. This mechanism might be related with the age of the \ion{H}{2} region,
(in general we expect younger regions to be denser); it can also be related to
shock waves (which would produce hot over-dense regions), that could be driven
by turbulence (of which more is expected in young regions).

The $P$(CLs)/$P$(RLs) derived from the temperatures and densities obtained 
from RLs by \citet{mna13} go from 1.32 to 530 with an average value of 89, if
we disregard NGC 3603 their average value becomes 34 a value more than one order of 
magnitude higher than the one derived by us. We consider that the overestimation
of the very weak \ion{O}{2} line intensities used by \citet{mna13}, that is partly due to
blends of these lines with even weaker lines, is the main reason for the 
differences in the derived temperatures and densities.

We decided to compare the radial velocities of the \ion{O}{2} and [\ion{O}{3}] 
lines to try to find out if there was any difference that could give 
us a clue on the study of thermal inhomogeneities. In Table \ref{Trad-vel} we 
present the median heliocentric radial velocity of five lines of 
the V1 multiplet of  \ion{O}{2}, lines for which their intensity decreases 
with increasing temperature, and compare them with those of the 
$\lambda$4363 [\ion{O}{3}] line, {which} originates in 
the O$^{++}$ region {and} whose intensity increases the most with temperature. 
We did not use the $\lambda$4959 and $\lambda$5007 [\ion{O}{3}] lines due
to two reasons: a) they are less temperature dependent than 
$\lambda$4363 and b) their shape might be affected by
 saturation effects. The average velocity difference between the 
 \ion{O}{2} and the  [\ion{O}{3}] lines of the sample amounts to 0.3 km/s, 
consistent within the uncertainties. This result is consistent with the 
idea that the \ion{H}{2} regions of the sample are chemically homogeneous.

\section{30 Doradus}\label{Sdoradus}

It is well known that the abundances derived from collisionally 
excited lines based on the 4363/5007 [\ion{O}{3}] temperatures, 
the so called direct method, are smaller than those derived from 
recombination lines{. This} difference has been known as the 
abundance discrepancy factor, ADF, the differences {for \ion{H}{2} regions} are 
typically of about a factor of 1.5 to 3. The ADF values pose 
two fundamental problems: a) which are the correct abundances, 
and b) which are the physical conditions responsible for the 
difference in the derived abundances.

We will use 30 Doradus to advance further on this problem. 
30 Doradus is a bright well observed \ion{H}{2} region that has 
most of its oxygen in the twice ionized state{. From} the UVES 
observations it amounts to 85\%, therefore its O/H 
ratio is one of the best studied among the observed galactic and 
extragalactic \ion{H}{2} regions. From the observations of 30 Doradus 
by \citet{pea03} there are at least eight qualitatively different 
determinations of the O/H ratio that can be obtained. In what follows 
we will present these determinations that exemplify some of the main 
methods that have been used to determine the O/H ratios in gaseous 
nebulae. We will discuss them in order of the derived O/H ratio.

The eight types of determinations that we will consider are:
1) Direct method (DM), 2) method based on the [\ion{O}{2}] and 
[\ion{O}{3}] nebular line intensities taking into account the degree 
of ionization and calibrated with DM abundances \citep{pil05} 
3) chemically homogeneous photoionization model 
\citep{tsa05}, 4) chemically inhomogeneous photoionization model 
\citep{tsa05}, 5) recombination lines method (RL), 
6) RL method plus the contribution to the O/H ratio due to the 
fraction of O tied up in dust grains, 7) method based on the intensity 
of the $\lambda$ 4363 [\ion{O}{3}] auroral line of a given object and the 
calibration by Pe\~na-Guerrero et al.(2012) based on RL abundances 
including the fraction of O tied up in dust grains, CALM method 
(Calibration based on the Auroral Lines Method), and 8) method 
based on the intensity of the  $\lambda$3727 [\ion{O}{2}] and  $\lambda$5007 
[\ion{O}{3}] nebular lines and the calibration by Pe\~na-Guerrero 
et al.(2012) based on \ion{O}{2} RL abundances including the fraction 
of O tied up in dust grains, RRM method (Revised $R_{23}$ Method).

In Table \ref{Tdor-O/H} we present the eight O/H determinations. The DM method 
is based on the assumption of $t^2 = 0.00$ and  the adoption of the 
temperatures derived from the ratio of the auroral and nebular lines 
of [\ion{O}{2}] and [\ion{O}{3}]. S1 is the homogeneous photoionization 
model computed by \citet{tsa05}, this model produces small temperature 
fluctuations that increase the O/H ratio by 0.03 dex relative to the 
value derived with the DM. Since the homogeneous photoionization model 
fails to reproduce the RLs of \ion{O}{2}, \citet{tsa05} presented 
an inhomogeneous  photoionization model with O rich low temperature 
clumps embedded into an \ion{H}{2} region with normal O abundances, 
the D2 model{. This} model adjusts properly many of the observed CLs an RLs 
intensities and leads them to two conclusions: a) the 
temperature variations could be explained by the presence of the O-rich 
(O/H = 9.30) low temperature clumps, and b) if this is the case, the overall abundance 
of 30 Doradus is intermediate between those derived from CLs and  RLs. 
The O/H gaseous abundance derived from the RL method is almost 
independent of the temperature, {and} the difference between the RL and DM 
methods is due to the temperature structure in the nebula. The RL 
plus dust method takes into account the fraction of O atoms tied up 
in dust grains \citep{pea10}. Finally the gas plus dust O/H ratios 
based on the CALM and RRM calibrations are also presented in 
Table \ref{Tdor-O/H}. These last two methods were calibrated using
determinations based on the RLs plus dust method, so we expect the last 
3 methods to agree within errors.

We consider that the best abundances for 30 Doradus, and for other 
\ion{H}{2} regions, are those given by the RL + dust method. If the 
RLs are not available for a given object, but the auroral lines are, 
the best determination is that given by CALM and if only the nebular 
lines are available the best determination is the one given by the RRM.

In Table \ref{Tdor-temp} we also include the $t^2$ values and the average $T_0$ 
values for the H$^+$ and O$^{++}$ zones predicted by the inhomogeneous 
photoionization model by \citet{tsa05} and the values derived from 
observations under the assumption of chemical homogeneity.

\section{Discussion}\label{Sdisc}

The ADF problem comes from trying to reconcile the abundances derived
from forbidden lines with those derived from recombination lines.  The
ADF problem is present in planetary nebulae and in \ion{H}{2}
regions. The ADF values in PNe can be due to four causes: temperature
variations, chemical inhomogeneities, strong density variations, and
non Maxwellian electron velocity distributions. For some objects it is
not easy to separate these causes, for example chemical
inhomogeneities produce temperature variations. For chemically
homogeneous nebulae the proper abundances are those given by RLs,
while in the presecnce of chemical inhomogeneities the representative
abundances are intermediate between RL and CL abundances.

Chemical inhomogeneities in some PNe are well established \citep
[e.g.][]{jac79,jac83,man88,haz80,tor90,liu00,bar06}, and probably they are the
dominant component of the ADF values higher than about 5, we consider
that most PNe with ADF values smaller than about 5 are probably
chemically homogeneous.  See the review by \citet{liu06} discussing
evidence in favor of chemical inhomogeneities in PNe, and the review
by \citet{pem06} discusssing evidence in favor of chemical homogeneity
for most PNe with ADF values smaller than about 5.

The effect of electron densities on the abundance determinations of
gaseous nebulae can also mimic spurious ADF
values. \citet{rub89,vie94,tsa11} have studied the dependence of the
line intensities on density when the upper energy levels producing
forbidden lines are de-excited by collisions. Depending on the line of
a given ion the critical density for collisional de-excitation is
different. If this effect is not taken into account the temperatures
derived from CLs are overestimated and the abundances are
underestimated. This can be the case for high density gaseous nebulae
and for certain ions. This effect is particularly relevant when
infrared lines are used to determine abundances and for objects of
relatively high density. We do not expect this effect to be important
for the objects studied in this paper, since we are mainly using the
4363 and 4959 [\ion{O}{3}] lines that have critical densities of $2.4
\times 10^7$ and $6.4 \times 10^5\ {\rm cm}^{-3}$ respectively, values
that are considerably higher than the densities of the \ion{H}{2}
regions considered here.

\citet{sta07} have discussed the possibility that the ADF might be due
to the presence of metal rich droplets inside \ion{H}{2} regions. Their 
model predicts that  the $n_d/n_{\rm H\,II}$ ratio is approximately ten, where
$n_d$ is the density in the droplets and $n_{\rm H\, II}$ is the density in the 
ambient \ion{H}{2} region. Since in this model most of the recombination 
line emission is expected to come from the metal-rich droplets and most 
of the collisionally excited line emission is expected to originate from
the ambient \ion{H}{2} region a ratio of
$n({\rm O \, {\small II}})/n\left<CL\right>$
 considerably higher than one is 
expected. From Table \ref{Tdensities} we obtain that for the UVES sample the average
$n({\rm O \, {\small II}})/n\left<CL\right>$ is approximatelly  0.5 
which is in conflict with the 
high metallicity droplets model.

\citet{tsa05} have presented a photoionization model with two
components, a component made of metal rich inclusions of low
temperature and high density in pressure equilibrium with the
other component. They adopt equal pressures for both components at
similar optical depths. For an increasing metallicity of the clumped
component its electron temperature decreases due to the more efficient
cooling from CELs, while because of the equal pressures, its electron
density proportionally increases.  The idea of taking
into account the behavior of the pressure in a two component model is
an excellent tool to study the possible presence of metal rich
inclusions of low temperature and high density. We will come back
to this idea further on.

\citet{nic12} have suggested that electrons do not have time to
thermalize in ionized nebulae so $\kappa$-distributions are better
suited than Maxwellian distributions to represent the electron
distributions in these objects. A $\kappa$-distribution can be
represented by a $t^2$ value, for objects with $\kappa > 10$ by the
following relation:
\begin{equation}
\label{Ekappa}
t^2 = 0.96/\kappa,
\end{equation}
this realtion is obtained from equations 16 and 18 of \citet{pem67}, and 
figure 10 of \citet{nic12}. The accuracy of the equatoin is better than 1\% 
in the 5000 to 20000 K temperature range.

We prefer the use of $t^2$ instead of $\kappa$ for the following
reasons: a) the $t^2$ formalism applies to many energy distributions
for the electrons, not only to $\kappa$-distributions; b) $t^2$ can be
used for objects with volume elements with different Maxwellian
distributions (different temperatures), as well as objects with volume
elements with different $\kappa$-distributions; c) to use
$\kappa$-distributions, the processes that produce these distributions
need to dominate the processes that produce Maxwellian
distributions, while the $t^2$ formalism can be used for nebulae with
Maxwellian distributions that include perturbations due to
$\kappa$-distributions, d) $\kappa$-distributions have an excess of
fast electrons (when compared with a single Maxwellian
distribution) but no excess of slow electrons, and thus can only
represent a limited number of physical processes, in particular they
can not represent physical mechanisms that produce cooler
regions, e.g. shadow ionization, while the $t^2$ formalism allows for
processes that produce hotter and/or cooler regions, e) $t^2$ values 
predicted by photoionized models make non negligible contributions 
to the $t^2$ observed values, these contributions cannot be fitted by 
$\kappa$-distributions.

\citet{pen12} present a list of 28 Galactic and extragalactic H II
regions with accurate $t^2$ determinations. The average ADF for this
set is 1.7, while the average $t^2$ value is 0.044. From Equation
\ref{Ekappa} this $t^2$ value corresponds to a $\kappa$ of 22.

The best \ion{O}{2} recombination lines to derive the electron
temperature and the electron density are those of multiplet 1, and the
best observations available are those of the UVES set, obtained with
the echelle of the VLT. The error in the temperature derived from the
$I$(4649)/$I$(4089) ratio of \ion{O}{2} is considerably higher than
the error in the temperature derived from the
$I$(4959,[\ion{O}{3}])/$I$(V1,\ion{O}{2}) ratio. The reasons are:
a)$\lambda$4089 is a very weak line {(moreover
other \ion{O}{2} lines that can be used to
obtain the $T_e$(\ion{O}{2}) value, e.g. 4189 and 4590, are expected
to be at least a factor of three weaker than 4089)}, b)$\lambda$ 4089 is blended with
other weak lines in particular with \ion{Si}{4} at $\lambda$4088.85,
c) the dependence of the temperature on the $I$(4649)/$I$(4089) ratio
is very {weak; we conclude from this discussion that higher signal to noise data to that of the UVES set is needed to derive the temperature based only on recombination lines}
. Alternatively: a) $\lambda$4959 is a very strong
line, b) $\lambda$4959 is not blended with lines of significant
intensity, c) the $I$(4959,[\ion{O}{3}])/$I$(V1,\ion{O}{2}) ratio
depends strongly on the temperature, d) in observations with the
quality of the UVES set the lines of multiplet V1 are not blended with
other lines, as can be seen in the figures of this multiplet presented
in the UVES papers and in the {agreement between the observed 
wavelengths with the theoretical ones.}

{
In this paper we derive an equation based on the new recombination
computations for \ion{O}{2} by Storey \citep[unpublished, but
available from:][]{liu12,mna13,fan13} to derive $T_e$ from the ratio
of [\ion{O}{3}] to \ion{O}{2} lines. We compare this temperature with
the $T_e$(4363/5007) temperature and obtain $t^2 ({\rm O}^{++})$ values that are in
very good agreement with the $t^2 ({\rm He}^{+})$ values derived from comparing of
$T_e$(\ion{He}{1}) to $T_e$(4363/4959), and with the $t^2 ({\rm He}^{+})$ values
derived from comparing the temperature determined from \ion{H}{1}
recombination lines and the Balmer continuum, $T_e({\rm Bac})$ to $T_e$(4363/5007). The
agreement among the three types of $t^2$ values implies that in these
objects H, He, and O are well mixed or, in other words, that these objects are chemically homogeneous and that there are no high density
low temperature knots inside the \ion{H}{2} regions of the UVES set.}
Moreover the $T_0({\rm O}^{++})$, $T_0({\rm He}^+)$, and $T_0({\rm
  H}^+$) values for a given object are similar. These results imply
that the \ion{H}{2} regions of our sample are chemically homogeneous
and that the ADF values are due to temperature variations,
that the O/H abundances derived from recombination lines of O and H
represent the correct O/H values. Other arguments in favor of the RL
abundances have been presented elsewhere \citep{pem11,sim11}.

{
From Figure \ref{Fne_OII} and the line intensities of the UVES set we 
derive the density of the \ion{H}{2} region based only on \ion{O}{2}
recombination lines and find that it is similar or lower than the
density derived from forbidden lines, this result also implies that
the \ion{O}{2} lines do not originate in high density knots.  Moreover
the electron presssure derived from the \ion{O}{2} densities is
similar to the electron pressure derived from the forbidden line
densities. The small pressure differences between the recombination
lines and the forbidden lines might be giving us clues about the cause
of the temperatue variations.}

{
Following the idea of the two components model one would expect the
pressure derived from the CLs to be similar to the pressure derived
from the RLs.
A study using the recombination lines of \ion{O}{2} and \ion{N}{2} to
determine temperatures and densities was made by \citet{mna13} for a
large number of PNe and \ion{H}{2} regions. In Table \ref{Tpres} we
present the ratio of the pressures from the \ion{O}{2} lines derived
from their Table 2 and those derived in this paper and find very large
differences. From the temperatures and densities derived by
\citet{mna13} we obtain an average P(CLs)/P(RLs) higher than 34 for
the nine \ion{H}{2} {regions} discussed in this paper (see Table
\ref{Tdensities}) . This result, if taken at face value, is contrary to the idea that
there are high density low temperature knots embedded in these gaseous
nebulae.} 

{
\citet{mna13} have in common with us five objects where they were able
to determine temperatures and densities from RLs of both N$^{++}$ 
and O$^{++}$, these results are presented in their Table 2.
The objects are M8, M17, NGC 3576, Orion, and 30 Doradus. From the 
ideal gas equation and the data presented by \citet{mna13} it is found that the 
\ion{N}{2} to \ion{O}{2} electron pressure ratio, $P$(\ion{N}{2})/$P$(\ion{O}{2}),
 presents a range of six orders of magnitude: from a ratio of 0.035 for the Orion nebula
to a ratio of 50,000 for 30 Doradus. Since we expect
the O$^{++}$ and N$^{++}$ regions to overlap substantially we expect
the measured electron pressure ratio to be close to one. We consider that a large
fraction of the range in the electron pressure ratio is due to large errors in
the measurement of the recombination line intensities.}

From the results of this paper, see Table \ref{Tdensities}, we obtain
an average for $P$(CLs)/$P$(RLs) for 8 \ion{H}{2} regions equal to
2.4. This result also indicates that there are no metal-rich,
high-density, low-temperature knots present in our sample.  Moreover
this result is significantly larger than 1.0 and might give us a clue
to explain the mechanism or mechanisms that produce the temperature
variations. This value implies that the regions where the CLs mainly
originate have slightly higher densities and temperatures than the
regions where the RLs mainly originate; these pressure differences
might be due to mild shocks or magnetic reconnection.

\section{Conclusions}\label{Sconc}

We present a set of equations and figures to derive $T$(V1/4959), 
$t^2$ (\ion{O}{2}), and $T_0$ (O$^{++}$), based only on forbidden and 
permitted lines of the O$^{++}$ region. 

The average temperatures and $t^2$ values derived from $T(4363/4959)$ 
and $T$(V1/4959) are in very good agreement with the $t^2$ values derived 
from H and He recombination lines. Moreover the $T_0({\rm O}^{++})$, 
$T_0({\rm He}^+)$, and $T_0({\rm H}^+$) values are similar. These results 
imply that the \ion{H}{2} regions of our sample are chemically homogeneous. 
Or in other words that H, He and O are well mixed in \ion{H}{2} regions,
that the ADF values are due to temperature variations, and that the O/H 
abundances derived from recombination lines of O and H represent the 
correct O/H values.

The densities derived from the recombination lines of multiplet 1 of 
\ion{O}{2} of our sample are in agreement {with} or smaller than the densities 
derived from the forbidden lines. The average pressure for the \ion{H}{2} 
regions of the sample derived from [\ion{O}{3}] collisionally excited lines 
is a factor of 2.4 higher than the pressure derived from \ion{O}{2} 
recombination lines. This difference might be significant{. If} this is the 
case it might be giving us information on the mechanism or mechanisms 
that produce the large observed $t^2$ values.

We present evidence against the presence of metal-rich,  temperature-low, 
density-high inclusions in the  \ion{H}{2} regions studied in this paper.

Of the several methods used to obtain the O/H ratio in \ion{H}{2} regions 
we consider the RL + dust to be the best one. The second best is the one 
based on auroral lines but corrected taking into account $t^2$, dust and 
the ionization structure, CALM method (Calibration based on the Auroral 
Lines Method). The third best is the one based on nebular lines but 
corrected taking into account $t^2$, dust and the ionization structure, 
RRM method (Revised R$_{23}$ Method). The direct method provides only a 
lower limit to the real O/H value.

We are grateful to an anonymous referee for a critical reading of the manuscript.
We are also grateful to Gary Ferland, Jorge Garc\'ia-Rojas, and Mar\'ia de los 
\'Angeles Pe\~na-Guerrero for fruitful discussions.
We received partial support from PAPIIT grant IN291129 and from
CONACyT grant 129753.

\appendix
\section{Line intensity ratios in the presence of temperature variations}\label{appendix}

\Hiirs\ have, in general, a temperature structure; for each ion present 
in the ionized region it is possible to use equations equivalent to 
equation \ref{Edef-T0} to determine $T_0\xion$, as well as 
equations equivalent to equations \ref{EI-e-W} and \ref{Edef-t2},
to derive $W\xion$ and $t^2\xion$.

When trying to estimate the intensity of a line $I(\lambda)$ as a
funcion of its emission coefficient in the presence of 
temperature variations, it is not possible to factorize
$\varepsilon(T_e)$ from integrals like the one in equation \ref{EI-e}: since 
$T_e$ is not constant, $\varepsilon(T_e)$ wont be either; the 
solution proposed by \citep{pem67} lies in expanding 
$\varepsilon(T_e)$ as a Taylor series arround $T_0\xion$:
\begin{equation}
\label{Ee-Taylor}
\varepsilon(T_e) = \varepsilon(T_0\xion) 
+ {(T_e-T_0\xion) \over 1!} {d \varepsilon \over dT}(T_0\xion) 
+ {(T_e-T_0\xion)^2 \over 2!} {d^2 \varepsilon \over dT^2}(T_0\xion) + ...
\end{equation}
from equations \ref{EI-e} and \ref{Ee-Taylor} we obtain:
\begin{eqnarray}
\label{EI-Taylor}
I= \int{\varepsilon(T_0\xion) n_e n\xion\over r^2} dV 
&+&\int{\varepsilon'(T_0\xion) (T_e-T_0\xion) n_e n\xion\over r^2} dV \nonumber \\
&+&{1 \over 2} \int{ \varepsilon''(T_0\xion) (T_e-T_0\xion)^2 n_e n\xion\over r^2} dV + ...
\end{eqnarray}
where $\varepsilon'$ and $\varepsilon''$ are the first and second 
derivatives of $\varepsilon$ with respect to $T$, and can be 
factored out of the integrals. The remaining integrals are related to 
$W\xion$, $T_0\xion$, and $t^2\xion$ respectively and equation 
\ref{EI-Taylor} can be written as:
\begin{equation}
\label{EI-W1}
I= \varepsilon(T_0\xion)W\xion +  
\varepsilon'(T_0\xion) (T_0\xion-T_0\xion) W\xion + 
{1 \over 2} \varepsilon''(T_0\xion) t^2\xion W\xion + ...
\end{equation}
The term asociated with $\varepsilon'$ will disappear since we chose 
to expand the Taylor series arround the average temperature, 
however the term associated with $\varepsilon''$ will not vanish; 
for moderate thermal inhomogeneities the contribution of the 
second order term will be much more important than higher order 
terms which we will ignore, therefore 
\begin{equation}
\label{EI-W1}
I= \varepsilon(T_0\xion)W\xion 
\left[1 + \left({T_0\xion^2\varepsilon''(T_0\xion) 
\over \varepsilon(T_0\xion)}\right) {t^2\xion \over 2} \right].
\end{equation}

When considering the ratio of the intensiteis of two lines 
originating from the same ion (the line intensities should have 
little or no density dependence), we obtain
\begin{equation}
\label{EI_ratio_1}
{I(\lambda 1) \over I(\lambda 2)}= 
{\varepsilon_1(T_0\xion) \over  \varepsilon_2(T_0\xion)}
\left[{1 + \left({T_0\xion^2\varepsilon''_1(T_0\xion) 
\over \varepsilon_1(T_0\xion)}\right) {t^2\xion \over 2} \over 
1 + \left({T_0\xion^2\varepsilon''_2(T_0\xion) 
\over \varepsilon_2(T_0\xion)}\right) {t^2\xion \over 2}}\right].
\end{equation}
In the regime of small temperature variations we obtain that
\begin{equation}
\label{EI_ratio_2}
{I(\lambda 1) \over I(\lambda 2)}= 
R_{\lambda 1 / \lambda 2}(T_0\xion)
\left[{1 + T_0\xion^2\left({\varepsilon''_1(T_0\xion) 
\over \varepsilon_1(T_0\xion)}
 - {\varepsilon''_2(T_0\xion) 
\over \varepsilon_2(T_0\xion)}\right) {t^2\xion \over 2}}\right].
\end{equation}

This shows that when using an uncorrected 
$R_{\lambda 1 / \lambda 2}(T)=\varepsilon_1 / \varepsilon_2$ 
to determine the temperature from the 
$I(\lambda 1) / I(\lambda 2)$ ratio, the determination will be 
skewed. To correct for the presence of temperature variations 
over the observed volume we  have that
\begin{eqnarray}
\label{EI_ratio_3}
{I(\lambda 1) \over I(\lambda 2)} = 
R_{\lambda 1 / \lambda 2}(T) 
&=&R_{\lambda 1 / \lambda 2}(T_0 + \Delta T) \nonumber \\
&\approx&R_{\lambda 1 / \lambda 2}(T_0) + 
\Delta T R'_{\lambda 1 / \lambda 2}(T_0) \nonumber \\
&\approx&R_{\lambda 1 / \lambda 2}(T_0) 
\left(1 + \Delta T {R'_{\lambda 1 / \lambda 2}(T_0\xion) 
\over R_{\lambda 1 / \lambda 2}(T_0\xion)}\right);
\end{eqnarray}
therefore
\begin{eqnarray}
\label{EDeltaT}
\Delta T {R'_{\lambda 1 / \lambda 2}(T_0\xion) 
\over R_{\lambda 1 / \lambda 2}(T_0\xion)} &=& 
\Delta T \left({\varepsilon'_1(T_0\xion) 
\over \varepsilon_1(T_0\xion)}
 - {\varepsilon'_2(T_0\xion) 
\over \varepsilon_2(T_0\xion)}
\right)  \nonumber \\
&=& T_0\xion^2\left({\varepsilon''_1(T_0\xion) 
\over \varepsilon_1(T_0\xion)}
 - {\varepsilon''_2(T_0\xion) 
\over \varepsilon_2(T_0\xion)}\right) {t^2\xion \over 2}
\end{eqnarray}
and finally
\begin{equation}
\label{E-T-t2}
T_{\lambda1/\lambda2}=T_0\xion 
\left[1 + T_0\xion\left({\,\,{\varepsilon''_1(T_0\xion) \over \varepsilon_1(T_0\xion)}
 - {\varepsilon''_2(T_0\xion) \over \varepsilon_2(T_0\xion)}\,\,
\over\,\, {\varepsilon'_1(T_0\xion) \over \varepsilon_1(T_0\xion)}
 - {\varepsilon'_2(T_0\xion) \over \varepsilon_2(T_0\xion)}\,\,}\right)
{t^2\xion \over 2}
\right];
\end{equation}
this equation relates the measured temperature to the average 
temperature $T_0$ and to the temperature variations parameter $t^2\xion$; observationally it 
means one measurement for two unknowns. If one wants 
to determine $T_0\xion$ and $t^2\xion$ one requires either two independent 
measurements of the temperature of a single ion (i.e. the 
measurement of at least 3 lines with different temperature 
dependence), or two determinations of temperatures for two ions
that occupy the same volume.

\begin{deluxetable}{lr@{$\pm$}lr@{$\pm$}lr@{$\pm$}lr@{}l}
\tablewidth{0pt}
\tablecaption{Temperatures from \ion{O}{2} and [\ion{O}{3}] lines
\label{Tt-OII}}
\tablehead{
\colhead{Object}&
\multicolumn{2}{c}{$T(4363/4959)$\TA}&
\multicolumn{2}{c}{$T({\rm V1}/4959)$\TB}&
\multicolumn{2}{c}{$T_0({\rm O}^{++})$\TB}&
\multicolumn{2}{c}{$T$(4649/4089)\TC}
}
\startdata
M16             & 7650 & 250 & 6295 & 135 & 6067 & 167 &  1000 & $^{+350}_{-460}$   \\
M8              & 8090 & 140 & 6756 &  84 & 6563 & 102 &  1400 & $^{+1600}_{-400}$  \\
M17             & 8020 & 170 & 6948 & 107 & 6805 & 127 &  4000 & $^{+6500}_{-1300}$ \\
M20             & 7800 & 300 & 6678 & 275 & 6513 & 328 & \multicolumn{2}{c}{$\le400$} \\
NGC 3576        & 8500 &  50 & 7238 &  60 & 7085 &  72 &  3160 & $^{+550}_{-300}$   \\
Orion           & 8300 &  40 & 7590 &  41 & 7518 &  47 & 15800 & $^{+2400}_{-2600}$ \\
NGC 3603        & 9060 & 200 & 7612 & 215 & 7462 & 251 &  7400 & $^{+10}_{-10}$     \\
S311            & 9000 & 200 & 7877 & 230 & 7777 & 261 & \multicolumn{2}{c}{$\le400$} \\
30 Doradus      & 9950 &  60 & 8902 & 104 & 8860 & 113 & \multicolumn{2}{c}{$\le400$} \\
\enddata
\tablenotetext{a}{Values from the VLT UVES papers.}
\tablenotetext{b}{This paper.}
\tablenotetext{c}{\citet{mna13} from \ion{O}{2} lines.}
\end{deluxetable}

\clearpage

\begin{deluxetable}{lr@{$\pm$}lr@{$\pm$}lr@{$\pm$}l}
\tablewidth{0pt}
\tablecaption{Other temperatures from the literature\TA
\label{Tt-lit}}
\tablehead{
\colhead{Object}&
\multicolumn{2}{c}{$T_0({\rm He}^+)$}&
\multicolumn{2}{c}{$T({\rm Bac})$}&
\multicolumn{2}{c}{$T_0({\rm H}^+$)}
}
\startdata
M16             & 7300 & 350 & 5450 &  820        & 5840 & 880 \\

M8              & 7650 & 200 & 7100 & 1100        & 7620 & 1180 \\
M17             & 7450 & 200 & 6500 & 1000\TB     & 6890 & 1060 \\
M20             & 7650 & 300 & 6000 &  300        & 6310 & 330 \\
NGC 3576        & 6800 & 400 & 6650 &  750        & 7110 & 800 \\
Orion           &      \mcnd & 7900 &  600        & 8290 & 630 \\
NGC 3603        & 8480 & 200 & 6900 & 1100\TB     & 7400 & 1180 \\
S311            & 8750 & 500 & \multicolumn{2}{c}{$\le10000$} & \multicolumn{2}{c}{$\le10700$} \\
30 Doradus      &      \mcnd & 9220 & 350         & 9640 & 370 \\
\enddata
\tablenotetext{a}{Values from the VLT UVES papers.}
\tablenotetext{b}{$T_e$(Paschen)}

\end{deluxetable}

\clearpage

\begin{deluxetable}{lr@{$\pm$}lr@{$\pm$}lr@{$\pm$}lr@{$\pm$}l}
\tablewidth{0pt}
\tablecaption{Temperature variations
\label{Tt-var}}
\tablehead{
\colhead{Object}&
\multicolumn{2}{c}{$t^2$(O$^{++}$)\TA}&
\multicolumn{2}{c}{$\left<t^2\right>$\TB}&
\multicolumn{2}{c}{$t^2$(\ion{He}{1}/CL)\TB}&
\multicolumn{2}{c}{$t^2$(\ion{H}{1}/CL)\TB}
}
\startdata
M16            & 0.042 & 0.007 & 0.039 & 0.006 & 0.017 & 0.013 &    0.045 & 0.014 \\
M8             & 0.041 & 0.008 & 0.040 & 0.004 & 0.046 & 0.009 &    0.022 & 0.015 \\
M17            & 0.033 & 0.005 & 0.033 & 0.005 & 0.027 & 0.014 &    0.035 & 0.021 \\
M20            & 0.035 & 0.012 & 0.029 & 0.007 & 0.017 & 0.010 &    0.049 & 0.019 \\
NGC 3576       & 0.039 & 0.003 & 0.038 & 0.009 &     \mcnd    &     0.037 & 0.017 \\
Orion          & 0.022 & 0.002 & 0.028 & 0.006 & 0.022 & 0.002 & $>$0.018 & 0.018 \\
NGC 3603       & 0.045 & 0.008 & 0.040 & 0.008 & 0.032 & 0.014 &    0.056 & 0.023 \\
S311           & 0.035 & 0.008 & 0.038 & 0.007 & 0.034 & 0.010 & $>$0.010 & 0.024 \\
30 Doradus     & 0.032 & 0.004 & 0.033 & 0.005 & 0.033 & 0.005 &    0.022 & 0.007 \\
\enddata
\tablenotetext{a}{This paper.}
\tablenotetext{b}{Values from the VLT UVES papers.}
\end{deluxetable}
\clearpage

\begin{deluxetable}{lr@{}lr@{$\pm$}lr@{}lr@{}l}
\tablewidth{0pt}
\tablecaption{Electron densities (cm$^{-3}$)
\label{Tdensities}}
\tablehead{
\colhead{Object}&
\multicolumn{2}{c}{$n_e$[\ion{Cl}{3}]\TA}&
\multicolumn{2}{c}{$n_e\left<{\rm CLs}\right>$\TA}&
\multicolumn{2}{c}{$n_e$(\ion{O}{2})\TB}&
\multicolumn{2}{c}{$n_e$(\ion{O}{2})\TC}
}
\startdata
M16        & 1370 & $\pm$1000            & 1120 & 220  & 660   & $\pm 230$   & 400 & $^{+3600}_{-160}$     \\
M8         & 2100 & $\pm$700             & 1800 & 350  & 560   & $\pm 130$   & 320 & $^{+100}_{-150}$      \\
M17        &\multicolumn{2}{c}{$\le630$} & 470  & 120  & 525   & $\pm 140$   & 200 & $^{+460}_{-100}$      \\
M20        & 350  & $^{+780}_{-350}$     & 270  &  60  & \nodata & \nodata   & \multicolumn{2}{c}{$\le100$}\\
NGC 3576   & 3500 & $\pm$800             & 2800 & 400  & 950   & $\pm 70$    & 560 & $^{+330}_{-140}$      \\
Orion      & 9400 & $\pm$1000            & 8900 & 200  & 3300  & $\pm 350$   & 3550 & $\pm$1000            \\
NGC 3603   & 5600 & $^{+3900}_{-2400}$   & 5150 & 750  & 1600  & $\pm 650$   & 220 & $^{+15}_{-40}$        \\
S311       &  \mcnd                      & 310  & 80   & 420   & $\pm 230$   & \multicolumn{2}{c}{$\le100$}\\
30 Doradus & 270 & $\pm$240              & 300  & 100  & 355   & $\pm 85$    & \multicolumn{2}{c}{$\le100$}\\
\enddata
\tablenotetext{a}{Values from the VLT UVES papers.}
\tablenotetext{b}{This paper.}
\tablenotetext{c}{\citet{mna13}.}
\end{deluxetable}

\clearpage

\begin{deluxetable}{lr@{}lr@{}l}
\tablewidth{0pt}
\tablecaption{Pressure ratios from CLs to RLs
\label{Tpres}}
\tablehead{
\colhead{Object}&
\multicolumn{2}{c}{$P$(CLs)\TA/$P$(RLs)\TA}&
\multicolumn{2}{c}{$P$(CLs)\TA/$P$(RLs)\TB}
}
\startdata
M16            & 2.14    & $\pm 0.86$        & 21.6 & $^{+22.0}_{-19.3}$ \\
M8             & 3.96    & $\pm 1.19$        & 32.5 & $^{+85.0}_{-24.0}$ \\
M17            & 1.06    & $\pm 0.39$        & 4.70 &  $^{+13.3}_{-3.3}$\\
M20            & \nodata &                   & \multicolumn{2}{c}{$\ge52.6$} \\
NGC 3576       & 3.54    & $\pm 0.57$        & 13.5 & $^{+6.0}_{-5.8}$ \\
Orion          & 2.98    & $\pm 0.33$        & 1.32 & $^{+0.80}_{-0.40}$ \\
NGC 3603       & 3.91    & $\pm 1.68$        & 530 & $^{+135}_{-46}$ \\
S311           & 0.85    & $\pm 0.52$        & \multicolumn{2}{c}{$\ge69.7$} \\
30 Doradus     & 0.95    & $\pm 0.38$        & \multicolumn{2}{c}{$\ge74.6$} \\
\enddata
\tablenotetext{a}{This paper }.
\tablenotetext{b}{\citet{mna13}.}
\end{deluxetable}

\clearpage

\begin{deluxetable}{lr@{$\pm$}lr@{$\pm$}lr@{$\pm$}l}
\tablewidth{0pt}
\tablecaption{Radial velocities\TA
\label{Trad-vel}}
\tablehead{
\colhead{Object}&
\multicolumn{2}{c}{(\ion{O}{2})\TB}&
\multicolumn{2}{c}{[\ion{O}{3}]\TC}&
\multicolumn{2}{c}{difference}
}
\startdata

M16         &  +4.3   &2.9    &  +3.4   &1.0    & -0.9    &3.1 \\
M8          &  -9.0   &2.2    &  -6.2   &1.0    & +2.8    &2.4 \\
M17         &  +7.6   &1.8    &  +5.5   &1.0    & -2.1    &2.1 \\
M20         &  +3.9   &2.7    &  -0.7   &1.0    & -4.6    &2.9 \\
NGC 3576    & -19.0   &3.0    & -16.9   &1.0    & +2.1    &3.2 \\
Orion       & +12.9   &1.2    & +14.4   &0.7    & +1.5    &1.4 \\
NGC 3603    & +15.1   &2.7    & +15.8   &1.0    & +0.7    &2.9 \\
S311        & +64.2   &2.3    & +66.0   &1.0    & +1.8    &2.5 \\
30 Doradus  &+231.1   &1.6    &+232.6   &0.7    & +1.5    &1.7 \\
\enddata
\tablenotetext{a}{Heliocentric radial velocities from the UVES set in km/sec.}
\tablenotetext{b}{$\lambda\lambda$4639,4642,4649,4651, and 4662 lines.}
\tablenotetext{c}{$\lambda$4363 line.}
\end{deluxetable}
\clearpage

\begin{deluxetable}{lcr@{$\pm$}lc}
\tablewidth{0pt}
\tablecaption{O/H Values for 30 Doradus
\label{Tdor-O/H}}
\tablehead{
\colhead{Model}&
\colhead{Components}&
\multicolumn{2}{c}{$12+log$(O/H)}&
\colhead{ADF}
}
\startdata
Obs., Direct Method\TA       &    gas    & 8.33 & 0.02       & 0.00     \\
Cal., Pagel's Method\TB      &    gas    & 8.33 &   0.05    & 0.00     \\ 
Mod., Homogeneous (S1)     &    gas    & 8.36 & 0.03      & 0.03     \\
Mod., Two-zone (D2)           &    gas    & 8.45 &  0.05     & 0.12     \\
Obs., Recombination Lines   &    gas    & 8.54 &  0.06     & 0.21     \\
Obs., RL+dust                     &gas+dust& 8.63 &  0.06     & 0.30     \\
Cal., CALM                          &gas+dust& 8.64 &  0.07     & 0.31     \\
Cal., RR$_{23}$M                &gas+dust& 8.68 &  0.08     & 0.35     \\
\enddata
\tablenotetext{a}{Direct method, $T$ assumed to be constant and given by the 
$I$(4363)/$I$(4959) ratio, consequently $t^2({\rm O}^{++})=0.000$ and 
$T_0({\rm O}^{++})=T(4363/4959)$.} 
\tablenotetext{b}{Pagel's Method is usually calibrated using the direct 
method.}
\end{deluxetable}
\clearpage

\begin{deluxetable}{lcccc}
\tablewidth{0pt}
\tablecaption{Average temperatures  and $t^2$ values for 30 Doradus
\label{Tdor-temp}}
\tablehead{
\colhead{Model}&
\colhead{$T_0$(H$^+$)}&
\colhead{$T_0$(O$^{++}$)}&
\colhead{$t^2$(Bac)}&
\colhead{$t^2$(O$^{++}$)}
}
\startdata
Obs., Direct Method\TA       &  \nodata     & $9950\pm60$ & \nodata         & 0.000               \\
Mod., Homogeneous (S1)\TB    & 9962         & 9818        & 0.0045          & 0.003               \\
Mod., Two-zone (D2)\TB       & 9654         & 8679        & 0.0223          & 0.078               \\
Obs., Recombination Lines\TA & 9640         & 9300        & $0.022\pm0.007$ & $0.038\pm0.005$     \\
\enddata
\tablenotetext{a}{\citet{pea03}.}
\tablenotetext{b}{\citet{tsa05}.}
\end{deluxetable}
\clearpage

\clearpage

\clearpage

\clearpage

\clearpage

\end{document}